 \documentclass[preprint]{aastex}

\shorttitle{Cluster Selection and Data Reduction}
\shortauthors{Kalirai, J. S. \it et al. \normalfont}

\begin{document}


\title{The CFHT Open Star Cluster Survey I -- Cluster Selection and Data Reduction}


\author{Jasonjot Singh Kalirai\altaffilmark{1}}
\affil{Physics \& Astronomy Department, 6224 Agricultural Road,
University of British Columbia, Vancouver, BC V6T-1Z1}
\email{jkalirai@physics.ubc.ca}

%\author{Harvey B. Richer, Gregory G. Fahlman, Patrick R. Durrell,
%Francesca D'Antona, and Gianni Marconi}

\author{Harvey B. Richer\altaffilmark{1}}
%\affil{University of British Columbia}
\author{Gregory G. Fahlman \altaffilmark{1,2}}
%\affil{Canada-France-Hawaii Telescope}
\author{Jean-Charles Cuillandre \altaffilmark{2}}
%\affil{Canada-France-Hawaii Telescope}
\author{Paolo Ventura \altaffilmark{3}}
%\affil{Osservatorio Astronomico di Roma}
\author{Francesca D'Antona \altaffilmark{3}}
%\affil{Osservatorio Astronomico di Roma}
\author{Emmanuel Bertin \altaffilmark{4}}
%\affil{Institut D'Astrophysique De Paris}
\author{Gianni Marconi \altaffilmark{5}}
%\affil{European Southern Observatory}

\and

\author{Patrick R. Durrell \altaffilmark{6}}
%\affil{Penn. State University}


\altaffiltext{1}{University of British Columbia}
\altaffiltext{2}{Canada-France-Hawaii Telescope Corporation}
\altaffiltext{3}{Osservatorio Astronomico di Roma}
\altaffiltext{4}{Institut D'Astrophysique De Paris}
\altaffiltext{5}{European Southern Observatory}
\altaffiltext{6}{Penn. State University}


\begin{abstract}
We present this paper in conjuction with the following as the
first results in the CFHT Open Star Cluster Survey.  This survey
is a large \it BVR \rm imaging data set of 19 open star clusters
in our Galaxy. This data set was taken with the CFH12K mosaic CCD
(42$'$ $\times$ 28$'$) and the majority of the clusters were
imaged under excellent photometric, sub-arcsecond seeing
conditions. The combination of multiple exposures extending to
deep (V $\sim$ 25) magnitudes with short ($\leq$ 10 second) frames
allows for studies ranging from faint white dwarf stars to bright
turn-off, variable, and red giant stars.  The primary aim of this
survey is to catalogue the white dwarf stars in these clusters and
establish observational constraints on the initial-final mass
relationship for these stars and the upper mass limit to white
dwarf production. Additionally, we hope to better determine the
properties of the clusters, such as age and distance, and also
test evolution and dynamical theories by analyzing luminosity and
mass functions. In order to more easily incorporate this data in
further studies, we have produced a catalogue of positions,
magnitudes, colors, and stellarity confidence for all stars in
each cluster of the survey. This reduction, along with the
computed calibration parameters for all three nights of the
observing run will encourage others to use this data in different
astrophysical studies outside of our goals. Additionally, the data
set is reduced using the new TERAPIX photometric reduction
package, PSFex, which is found to compare well to other packages.

    This paper is intended both as a source for the astronomical
community to obtain information on the clusters in the survey and
as a detailed reference of reduction procedures for further
publications of individual clusters. We discuss the methods
employed to reduce the data and compute the photometric catalogue.
We reserve both the scientific results for each individual cluster
and global results from the study of the entire survey for future
publications.  The first of these further publications is devoted
to the old rich open star cluster, NGC 6819, and immediately
follows this paper.

\end{abstract}


\keywords{color-magnitude diagrams -- open clusters and
associations: general -- methods: data analysis -- techniques:
photometric -- white dwarfs}


\section{Introduction}

    The major effort in star cluster photometry has been directed
towards globular clusters as they are some of the oldest
structures in the Universe and can set a lower bound for the age
of the Universe.  These clusters were formed many billions of
years ago and contain red, metal poor stars (Population II).
Globular clusters can contain up to a million stars whereas the
richest open clusters contain several thousands of stars at most.
The stars in these less populated clusters are much bluer,
younger, and metal rich (Population I) than the globular cluster
stars.  An in-depth analysis of these less crowded open star
clusters is crucial to provide insight into the less well
established theoretical stellar evolution models available for
such clusters.

    The present survey is a large \it BVR \rm imaging data set of 19
open star clusters in our Galaxy. This data set was taken with the
CFH12K mosaic CCD (42$'$ $\times$ 28$'$) and the majority of the
clusters were imaged in excellent photometric, sub-arcsecond
seeing conditions.  There are very few other programs currently
underway to address these issues, partly because a large field of
view and a large telescope are required.  Many papers have been
published on individual open star clusters (some of which are in
this survey), but the only major current survey similar to ours is
the WIYN Open Star Cluster Survey (WOCS). There are a few
overlapping clusters in these two surveys however some of the
science goals are different. The focus of the present study will
not only include the measurement of key properties for most of the
star clusters (such as age and distance), but will also test
theoretical models by fitting isochrones.  The ultimate goal of
our survey is to catalogue the white dwarf stars for each cluster
and constrain both the initial-final mass relationship for these
stars and the upper mass limit to white dwarf production, both of
which are currently rather poorly constrained by observational
data (see \cite{weidemann} for a review).

    The CFHT data for these clusters is unprecedented with
regard to the diversity of the sample, the size of the data set,
and the precision of the measurements. The sample of clusters,
which span a large range of angular sizes, age, and metallicity
values, were chosen on the basis of stellar density, age, and
distance. Previous photometry on these clusters has been mostly
limited to photoelectric and photographic observations, and
usually concentrated on bright to intermediate magnitude ranges.
The color-magnitude diagrams from these observations generally
show a large amount of scatter both in the main-sequence and
turn-off stars.  The faintest stars recorded in most studies are
$\sim$17-18th visual magnitude.  While, the present photometry
includes the bright end of the main sequence, the program is
primarily intended to find white dwarf stars down to V = 25, and
therefore provides some of the deepest images ever taken for open
star clusters.  A very well defined main sequence should be seen
over a long magnitude range (10-12 magnitudes) for the rich, young
clusters, which will allow for a wide range of investigations
including main sequence termination, model fitting, and luminosity
and mass functions.  A small number of the clusters are very sparsely 
populated, but also among the youngest in the survey.  These 
are especially important for white dwarf searches because any white 
dwarfs found in these clusters would have had to form from very 
massive progenitors, and therefore they will establish a constraint 
on the upper mass limit to white dwarf production.

    Theoretical isochrones are commonly fit to observational results
in order to validate the theory and/or determine the properties of
the cluster. The lower main sequence often contains more scatter
and errors due to photometric uncertainty for low signal to noise
stars, so these isochrones have commonly been fit to the upper
main sequence and turn-off only. The current data set goes fainter
(V $\sim$ 25) than the termination point of the main sequence for
some of the clusters and will provide an excellent test to
theoretical isochrones on the lower main sequence (10 $\leq$
M$_{V}$ $\leq$ 12). Of particular interest is the change in slope
of the main sequence caused by the various stages of stellar
evolution and structure, such as the onset of H$_{\rm2}$
dissociation-recombination in the stellar envelope at
(B$\rm-$V)$_{0}$ = 1.0$\rm-$1.1, or the `kink' in the main
sequence at (B$\rm-$V)$_{0}$ = 0.3$\rm-$0.4 caused by changes from
convective envelope models to radiative models. These slope
changes will be used as guidelines for the fit of the observations
with the theoretical models. The survey will use new up-to-date
models calculated by the group at the Rome Observatory (Ventura \&
D'Antona, Private Communication).

    In addition to model testing, there are many other scientific
goals which we hope to complete with this data set.  For example,
a main sequence turn-on could potentially provide an independent
age measurement for some of the very young clusters for which the
low mass stars have not yet reached the main-sequence.  The faint
magnitudes reached by such a study will also allow the study of
white dwarf stars.  Fitting models to the end of the white dwarf
cooling sequence, if bright enough, will allow a third
determination of the age measurement of some of the youngest
clusters.  For example, we use white dwarf cooling models
\citep{richer} to determine that the white dwarf cooling sequence
in an intermediate age cluster ($\sim$250 Myrs) will terminate at
M$_{V}$ $\sim$ 11.2, depending on the mass of the stars, whereas
for a much older cluster ($\sim$2 Gyrs), the end of the cooling
sequence occurs at a significantly fainter magnitude of M$_{V}$
$\sim$ 13.5. Therefore, even for a moderately close cluster, the
depth of the current photometry should allow us to establish the
white dwarf cooling age for a large number of the clusters in the
survey (see Table 1).  For the older clusters for which the
termination of the white dwarf cooling curve is fainter than our
limiting magnitude, we should nevertheless be able to identify a
significant number of white dwarf candidates above our mean cut
off of V $\sim$ 25. Other studies that will be looked at involve
producing luminosity and mass functions for these clusters.  In
particular, these will be explored for evidence of dynamical
evolution in the older clusters.  An updated mass-luminosity
relation from new stellar models will also be used to test mass
segregation in the clusters by comparing the mass functions for
various annuli at different distances from the centers of the
clusters. This test will be possible for a wide range of masses
($\sim$0.5 $\rm-$ 4M$_\odot$) for most of the intermediate aged
clusters. Improved distance estimates of the clusters will be
found by establishing a fiducial of the main sequence and
comparing to the Hyades main-sequence which has a very well
established distance from Hipparcos measurements \citep{perryman}.
Chemical evolution theories for old stars in the clusters will be
tested by comparing the masses of white dwarfs to their
progenitors on the main-sequence. Number counts involving the red
giant/white dwarf ratio will be carried out by assuming
conservation of star number through various stages during stellar
evolution, to establish the number of expected white dwarfs.
Binary star tests will be undertaken to determine the population
of binaries (relative to main-sequence stars) in each cluster.
These have been shown to have important evolutionary effects,
especially for the younger, less dense clusters
\citep{delafuentemarcos2}.

Additionally, since most of these clusters are confined to the
disk of the Galaxy, this work allows for a better understanding of
the dynamical evolution of the Galactic disk as well as the
Galactic star formation history.

Our photometric catalogue has been produced in order to encourage
and expedite studies outside the scope of our goals.  The current
paper will introduce these clusters and the specific photometry
for each one. Answering the questions raised above will be
reserved for a cluster-by-cluster analysis and will be presented
in subsequent papers as a part of the CFHT Open Star Cluster
Survey. The first of these papers, entitled `The CFHT Open Star
Cluster Survey II : Deep CCD Photometry of the Old Open Star
Cluster NGC 6819', immediately follows this paper.

\section{Observations}\label{observations}

    The data for these 19 clusters (see Table 1) was taken during
an excellent three night observing run with the
Canada-France-Hawaii Telescope (CFHT) on October 15-18th 1999,
using the new CFH12K camera. The optical detector is a 12,288 by
8,192 pixel CCD mosaic camera for high resolution wide field
imaging at the CFHT prime focus. This mosaic camera is the largest
close-packed CCD currently being used for astronomical research.
The CCD camera was ideal for our purposes as it covers a large
area on the sky (42 by 28 arc-minutes, or about 1.5 times the area
of the full moon), and also contains a large number of pixels
($\geq$ 10$^{8}$) to ensure a high angular sampling.  The camera
is equipped with twelve 2048$\times$4096 pixel CCDs with an
angular size of 0.206 arc-seconds per pixel at the f/4 prime
focus.  The orientation of the CCDs within the mosaic is such that
chips 00-05 form a sequence on the bottom row from left
$\rightarrow$ right, and chips 06-11 are directly on top of the
bottom row.  Therefore the inner 4 CCDs are chips 02, 03, 08 and
09, and the outer 4 CCDs are chips 00, 05, 06 and 11.

    Data for each cluster was taken in three filters: B, V and R.
The R images were usually rather shallow and are used to provide
some leverage for the reddening of the clusters based on
color-color diagrams.  The optimal exposure time to be used was
determined so as to achieve a limiting magnitude 1 magnitude
fainter than the oldest expected white dwarfs in most of the
clusters.  This was found by equating the cluster age to the white
dwarf cooling time \citep{richer}.  For a few of the older
clusters (such as NGC 6819), it was not feasible to try and
observe the end of the white dwarf cooling sequence from ground
based imaging as it was expected to be too faint, so we hope to
detect as many white dwarfs as possible above our limiting
magnitude.

     To achieve a higher signal to noise ratio, multiple 300 second
exposures were taken for some of the clusters in both the V and B
filters.  Additionally, single 50 and 10 second exposures were
obtained in all three of the B, V, R filters.  The long exposures
in each band were dithered from one another to prevent a star from
landing on a bad pixel in more than one frame.  The multiple
fields were averaged and combined together using the FITS Large
Images Processing Software (FLIPS) (see section \ref{FLIPS}). In
many cases the CFH12K images for these clusters are the deepest
yet, and indicate a much richer cluster population than previously
thought.  The images also provide a more complete aerial coverage
for each of the clusters than most previous studies. In addition
to these frames, several flat-field images and bias frames were
also taken (see section \ref{datared}).  Five of the clusters in
our sample have sizes that exceed that of the CCD mosaic field of
view, and thus additional images of neighboring blank fields were
also taken. These will be used for background subtraction by
obtaining an estimate of the number of field stars in the region
around the cluster. Table 2 summarizes the observational data that
was taken for each cluster in the survey, as well as other
relevant information pertaining to the exposures.

    Since most of these clusters have not been extensively studied
in the past it is difficult to compare results with `good'
previous photometry. Furthermore, none of the clusters contain any
well established standard stars in the field.  These factors lead
to the critical calibration stage of the reduction which we
address by taking multiple images of the SA-92 and SA-95 standard
star fields \citep{landolt}.  Calibration is further discussed in
section \ref{calib}, where we provide details of numbers of
stars/chip used to calibrate the data set (see Tables 2 \& 3 for
particulars).

%% In this section, we use  the \subsection command to set off
%% a subsection.  \footnote is used to insert a footnote to the text.


%% This section also includes several of the displayed math environments
%% mentioned in the Author Guide.

\section{Data Reduction}\label{datared}

    The data for the survey was reduced and organized
locally at CFHT.  The first stages of the reduction involve
pre-processing the raw data as shown in equation
(\ref{flatdebias}). First several zero exposure bias frames are
taken and subtracted from the images (de-biasing) to account for
counts read out even if no light falls on the CCD.  A dark current
is also subtracted from the longer exposures. After the images are
de-biased, the data is flat-fielded to account for pixel-to-pixel
variations. For the flat-field frames, we combined twilight flats
taken from all three nights of the observing run.
%-----------------------------
\begin{equation}
Preprocessed\ Frame = \frac{raw\ image - bias}{flat - bias}
\label{flatdebias}
\end{equation}
%-----------------------------

\subsection{FLIPS} \label{FLIPS}
\subsubsection{Pre-processing} \label{pre}

    FITS Large Images Processing Software (FLIPS) is a
highly automated software package developed at CFHT (Cuillandre
2001).  FLIPS was originally developed for early CCD mosaic wide
field imagers at CFHT, such as MOCAM (4k $\times$ 4k pixels, 1994)
and UH8k (8k $\times$ 8k pixels, 1995), but has now been updated
and upgraded with new functions to handle CFH12K images.  This
software is ideal for our survey as it is optimized for both speed
and requires limited memory resources.  FLIPS is not a photometric
reduction package and should be thought of as a package which
performs similar functions to the commonly used \sc iraf \rm task
\bf mscred\rm.

    FLIPS is designed to operate on individual chips
within the CFH12K mosaic.  The first steps involve averaging the
`good' exposures for each of the bias, dark and flat exposures.
For each filter, these exposures can be accepted or rejected based
on the on-screen statistics (ie. if image level is too high
(saturation) or too low (no flux)). The flats used in the survey
are twilight exposures averaged and sigma clipped (iteratively 
eliminating if $\pm$sigma cut is not satisfied) from all three
nights of the run, whereas the darks were combined by taking a
median from a sigma clipped sample. FLIPS will then search through
all the images for each cluster and apply the above corrections
(bias, dark, flat) based on input parameters specific to the CCD
and data. These detailed parameters for each of the operations
have been optimized to produce the best processing. One of the
important features of FLIPS is that it normalizes the background
sky value to the chip with the highest sky value (lowest gain),
CCD 04. This provides for a scaled data set with a smooth
background on all chips. Therefore, the instrumental zero points
for the data set will all be almost equal.  The final processed
data show very small variations from a completely smooth
background.  By measuring the statistics in small boxes of size
comparable to the mean stellar aperture size at various positions
in the mosaic, we find the flat-fielding to be good to $\sim$0.5\%
in V and $\sim$0.7\% in B, averaged over 11$''$ sq. patches.

\subsubsection{Combining Multiple Exposures} \label{combine}

    In order to achieve a higher signal to noise ratio and a deeper
color-magnitude diagram, we obtained several deep exposures for
some clusters.  These exposures were dithered from one another
(see section \ref{observations}) and then aligned and combined
into one image. We use a program called \bf align\rm, within
FLIPS, which searches through a specified area on each CCD and
finds patterns of common stars based on both the positions and
fluxes of the stars. This pattern recognization algorithm works
very quickly for moderately rich clusters, however it is quite
slow for sparse fields. FLIPS \bf align \rm uses SExtractor
\citep{bertin} to create an X Y catalogue of aperture magnitudes 
and positions which \bf align \rm searches
through to find the geometric patterns.  This catalogue does not
not contain the final photometry as these magnitudes are only used
for this one purpose.  An important aspect of \bf align\rm, as we
will shortly see, is that it defines the true sky background and
atmospheric transmission across the data set. Next we invoke the
\bf imcombred \rm command within FLIPS to register the frames with
respect to one common image, and average the data.  The parameters
for the variable transmission (if present) and sky background are
taken and applied directly from the \bf align \rm output. \bf
Imcombred \rm re-centers the individual images based on the X \& Y
offsets derived for each CCD.  For a given pixel coordinate, a
column of pixels from each of the individual images is created
with the proper background and transmission corrections.  Then a
CCD sigma clipping of $\pm$5 is applied on that set of pixels to
determine the exact shifts. This requires a determination of the
readout noise and gain to evaluate properly the expected noise in
the signal and reject outliers.

    We found that the point-spread-function fit to the stars was being
skewed in the average because the centers of stellar positions
were not being located accurately enough.  In order to correct the
problem, we use a FLIPS \bf imcombred \rm option called
sub-pixelling.  This sub-pixelling parameter can be given a value
of n = 1, 3 or 5, and will split each CFH12K pixel (0$\farcs$206)
into n$^{2}$ boxes, and then locate the center within each of
these sub-divisions. This procedure is similar to the \bf drizzle
\rm technique in \sc iraf \rm and works to better align the
stellar profiles directly on top of one another in the above
combining procedures.  The sub-pixelling feature within FLIPS also
maintains the original resolution of the image and does not lose
any spatial information in the combining process. The resulting
CMD (for n = 5) was also found to go about 0.3 magnitudes fainter,
thereby indicating that the additional step allowed for the
measurement of lower signal to noise objects. This division led to
a very slow computing process of averaging the frames (as expected
for 2.5 billion pixels on each mosaic), however the results were
excellent.

The resulting data images are next used in the TERAPIX
PSF-extraction and modelling tool, PSFex, to find magnitudes,
colors, errors, and stellarity.

\section{PSFex}

     Point-Spread-Function Extractor (PSFex) is a highly
automated program which will be integrated in the code of the next
SExtractor version (Bertin, Private Communication).  SExtractor
detects sources through a segmentation process consisting of 6
essential steps: estimation of the sky background, thresholding,
deblending of overlapping images, filtering of the detections,
photometry, and star/galaxy separation.  Further details and
simulation results are given in \cite{bertin}.  SExtractor is most
commonly used to distinguish between stars and galaxies at faint
magnitudes by assigning a stellarity index to the objects. This
process also eliminates bad pixels caused by cosmic ray hits. The
stellarity is key to our project as white dwarf stars and faint
star forming galaxies both appear as faint, blue objects.
SExtractor determines the stellarity of objects by computing a
neural network (a group of connected units) which learns based on
other (high signal to noise) stars in the field. The
classification is highly dependent on the seeing of the image and
generally will be less accurate for faint stars than faint
galaxies because of crowding: faint stars have a higher
probability of catching wings added to the profile by a background
galaxy which would result in misclassification. The shape of all
diagrams (stellarity vs. magnitude) that we produced using
SExtractor agree with that of the Monte Carlo experiments executed
during the tests of the package by Emmanuel Bertin. Furthermore,
our results seem to indicate that a simple constant magnitude
stellarity cut (0.95 is commonly used) may not be a real
indication of the separation of stars from galaxies.  We find a
common pattern in the stellarity vs. magnitude diagram for all
rich clusters, and this may be a more accurate indication of the
separation between the galactic and stellar sequences.  The above
points are illustrated in Figure \ref{N2099stellb}, for the stars
of the young, rich open cluster NGC 2099.  This type of cut would have 
to be used with caution however as their are some objects that fall 
under the line, yet are brighter and have a slightly higher stellarity 
than sources that are over the line. More data will eventually allow us 
to decide on an optimal stellarity cut. 

     PSFex automatically creates a PSF based on a set of bright,
unsaturated and isolated stars.  Included in the PSF are
polynomial basis functions which can map the variations of the PSF
across the field. This is done immediately after the SExtractor
find catalogue is created. Next, it uses this PSF to derive PSF
magnitudes and colors for the objects in the original catalogue.
This entire process is not very computer intensive and is easily
executed with very little user interaction.

    For the CFHT Open Star Cluster Survey, we have chosen to use a
variable PSF for all fits to account for small changes in the
profiles of stars over the large range of each CCD. Additionally,
there are differences between each of the 12 CCDs on the mosaic so
the analysis is done on a single chip basis for all exposures.
Preliminary results, based on the differences in measured
magnitudes of PSFex and ALLSTAR \citep{stetson}, the shape of the
main-sequence in PSFex compared to ALLSTAR, and the number of
stars that were measured indicate consistency between the two
packages.  To better illustrate the small scatter measured for
common stars in these two packages, we show a comparison of the
difference in magnitudes vs magnitude for the two packages in
Figure \ref{comp}.  Results on other CCDs and other frames were all
much better than this, some by several factors.  A photometric error 
bar (combining errors from both programs) is also shown in order to 
judge the agreement between the two systems. The spread (2/3$\sigma$) 
is within the errors along the tight vertical sequence of stars, 
except at the very bright and faint ends.  The scatter at the bright 
end is in part due to saturation and also due to the handling of blended 
objects in the frames.  Although the low end scatter may also appear 
to be larger than the errors this may be simply due to different 
detection parameters between the two packages.  Our expected limiting 
magnitude is indicated on Figure \ref{comp}.  The results for different 
magnitude cuts, in the form of a histogram, are also shown in Figure 
\ref{histogram}.  The distribution peaks strongly at $\Delta$B = 0 and 
quickly falls off indicating a very small amount of scatter in all but 
the very faint case (see lower right diagram).

\section{Calibration of Instrumental Data} \label{calib}

\subsection{Landolt Standard Stars} \label{landolt}

  Observations of Landolt (1992) standard stars in SA-92 and SA-95
were obtained in order to convert the instrumental magnitudes to
real magnitudes as shown in equations (\ref{vshift}) and
(\ref{bshift}).  We use a total of 23 calibration frames of
different exposure times, air-masses, and wavelength filters to
transform the data from instrumental magnitudes to calibrated
magnitudes. Table 2 provides information on the number of
exposures for each frame, on each night.

%-----------------------------
\begin{equation}
v_{instr} = V + \alpha X + \beta (B-V) + Z_{v}
 \label{vshift}
\end{equation}
%-----------------------------

%-----------------------------
\begin{equation}
b_{instr} = B + \alpha'X + \beta' (B-V) + Z_{b} \label{bshift}
\end{equation}
%-----------------------------

In these equations $v_{instr}$ and $b_{instr}$ are 
instrumental magnitudes (discussed more later).  For equation
(\ref{vshift}), $\alpha$ is the coefficient of the air-mass term
X, $\beta$ is the coefficient of the color correction term
(B$\rm-$V), and $Z_{v}$ is the zero point shift for the V-band.
Similarly, in equation (\ref{bshift}), $b_{instr}$, $\alpha'$,
$\beta'$, and $Z_{b}$ are the corresponding parameters for the
B-band images.  There are systematic differences between chips on
the CCD so separate calibrations are required for each chip. The
color term does not change with different exposures or nights of
observations. The zero points are the most critical part of this
first stage of the calibration equations. This shift tells us the
relationship between the instrumental magnitudes and the real
magnitudes, taking the exposure time into account.

    The magnitudes of the standard stars used by Landolt are based
on non-linear transformations of the magnitudes and colors of the
stars in several different filter bands.  The final magnitudes
derived for the stars are based on the total amount of flux that
the star emits.  However, the fitting parameter used in our PSF
photometry is one that maximizes the signal to noise ratio. In
order to collect the entire light from the star without an
increase in the background level, we use aperture photometry for
calibration. For those stars which are not saturated and appear in
both Landolt's paper \citep{landolt} and our images of SA-92 and
SA-95, we use \sc daophot \normalfont to produce a curve-of-growth
of flux for aperture sizes ranging from 8 to 20 pixels.  At an
aperture size of 16 pixels it is found that the magnitude begins
to level off (less than 0.01 magnitude difference in all cases).
Therefore, an aperture size of 16 pixels collects almost all of
the light from the star ($\geq$ 99\%). For these stars we use the
flux from this aperture in our calibration equations.

    To prevent Landolt (1992) objects from being saturated on our
images, we choose extremely short exposure times depending on the
filter.  The number of exposures that were used in the calibration
for each night is summarized in Table 3.

    The calibration parameters are found by using a least squares
method to determine the coefficients of the air-mass, color and
zero point terms.   The specific algorithm is based on a method
designed by Harris \citep{harris}. These parameters are solved for
in those chips which contained enough calibration stars for the
calibration program to produce accurate (low sigma) results. The
chips with only a few or no calibration stars do not help this
analysis. Some sections of Landolt's frames were highly
concentrated with calibration stars however, the CFHT chips did
not align with these regions when the calibration frames were
taken.  It would be strongly preferred to have enough calibration
stars in each chip to get accurate parameter values, which could
then be used individually. Additionally, dithering the calibration
frames could have allowed for the measurement of common standards
on different chips, which would also help in better constraining
individual CCD calibration parameters. We are forced to average
the `good' chips (02,04,07,08) together and use the resulting
parameters on other chips.  There is minimal scatter in the
parameters for different chips so this method is reliable.  For
example, the photometric uncertainty in the zero points for the
standard stars on all four chips during night 1 of the observing
run were measured to be $\sim$0.015 in V and $\sim$0.014 in B. The
air-mass coefficients were determined to be 0.088 $\pm$ 0.01 in V
and 0.165 $\pm$ 0.005 in B, both in good agreement with CFHT
estimations of 0.10 and 0.17 respectively. The color terms were
averaged over the three night observing run and are in agreement
with CFHT estimations in the V filter and slightly lower than
estimations for the B filter.  Similar estimates are also obtained
for night 2, whereas for night 3 there are too few observed
standards to establish low sigma calibration parameters.
Therefore, we use the transformation equations from night 1 and 2
to create more standards and then apply these to the night 3 data.
These results were also within the errors given above.

    To illustrate the small scatter between the actual magnitude of
each standard star, and the calculated value from the least
squares solution, we present a composite O-C diagram in Figure
\ref{calibfigure}. There is clearly no obvious bias in this
diagram, which includes all calibration stars on all nights, in
all bands. O-C diagrams for data in each of the separate V and B
filters, individual nights of the observing run, or red vs. blue
segregation also shows no trends or biasses.  The small vertical
spread is caused by measurements of the same star at various
air-masses and/or exposure times.

\subsection{On-Field Standards} \label{onfield}

    As mentioned earlier, we have determined all of the best fit
calibration parameters in both the V and B filters.  One can now
simply use these in equations (\ref{vshift}) and (\ref{bshift}) to
determine a preliminary zero point term.  Next, we find $Z'_{v}$
and $Z'_{b}$, which are denoted in equations (\ref{vcal}) and
(\ref{bcal}) as the final shifts of the instrumental magnitudes to
the calibrated magnitudes, by creating on-field standards on the
science frames.  These new zero points will include the correction 
needed in going from aperture to PSF magnitudes. The 
stars that are used for this are isolated and bright, yet
non-saturated. Finally, equations (\ref{vshift}) and
(\ref{bshift}) are solved for $V + \beta (B-V)$ and $B + \beta'
(B-V)$ which are then used in equations (\ref{vcal}) and
(\ref{bcal}), along with the PSF magnitude values to derive
$Z'_{v}$ and $Z'_{b}$ for each chip. These new zero points can be
used to calibrate the entire data set to calibrated magnitudes as
they are particular to the science frames only. This shift is
employed by first solving the calibration equations for a color
term, as shown in equation (\ref{B-V}), and then applying the
color term to equations (\ref{vcal}) and (\ref{bcal}) to get the V
and B magnitudes. This method can be used for all clusters in the
survey by using the determined calibration parameters for the
night in which the respective cluster was imaged.

%-----------------------------
\begin{equation}
V = v_{PSF} - Z'_{v} - \beta (B-V)
 \label{vcal}
\end{equation}
%-----------------------------

%-----------------------------
\begin{equation}
B = b_{PSF} - Z'_{b} - \beta' (B-V) \label{bcal}
\end{equation}
%-----------------------------

%-----------------------------
\begin{equation}
B-V = (1+(\beta'-\beta))^{-1}[(b-v)_{PSF}-(Z'_{b}-Z'_{v})]
\label{B-V}
\end{equation}
%-----------------------------

\section{Results} \label{results}

At the current time, we have completed reductions for two clusters
in the survey, NGC 2099 and NGC 6819, and have begun reducing two
additional clusters. These clusters are both relatively rich in
stellar content, though they are quite different in age. Following
this paper we present a separate paper of complete results on NGC
6819 \citep{kalirai}. This will shortly be followed with the
analysis of NGC 2099, at which time we will start evaluating the
clusters based on a pre-determined priority system. The results
for NGC 2099 and these other clusters will include white dwarf
star analysis, white dwarf cooling ages, isochrone fitting,
statistical subtraction of field stars, luminosity and mass
functions, and binary star studies.

\section{Conclusion} \label{conclusion}

    We have determined the instrumental magnitudes for the entire
CFHT Open Star Cluster Survey.  Soon, we will make this
photometric catalogue available for general studies.  The
catalogue will consist of positions, magnitudes, colors, and
stellarity for the 19 open star clusters in the survey.
Preliminary results and color-magnitude diagrams for four of the
star clusters that we have looked at indicate a strong potential
for successful completion of the global goals of the survey: \\

1.) Constraining the initial-final mass relationship for white
dwarf stars in these clusters.

2.) Establishing observational constraints to the upper mass limit
to white dwarf production.

3.) Refining parameters, such as age and distance, for the
clusters.

4.)  Testing up-to-date theoretical models.

5.)  Investigating evolutionary processes in the cluster by
exploring the luminosity and mass functions.

The rich stellar content of these clusters, large aerial coverage,
and the excellent photometry, allows for a large range of
astronomical studies outside of our goals: the study of bright
variable stars in open cluster, core-overshooting constraints for
turn-off stars, astrometric studies etc... The results for NGC
6819, the first detailed studied cluster in the survey follow in
the next paper.

\clearpage

\figcaption[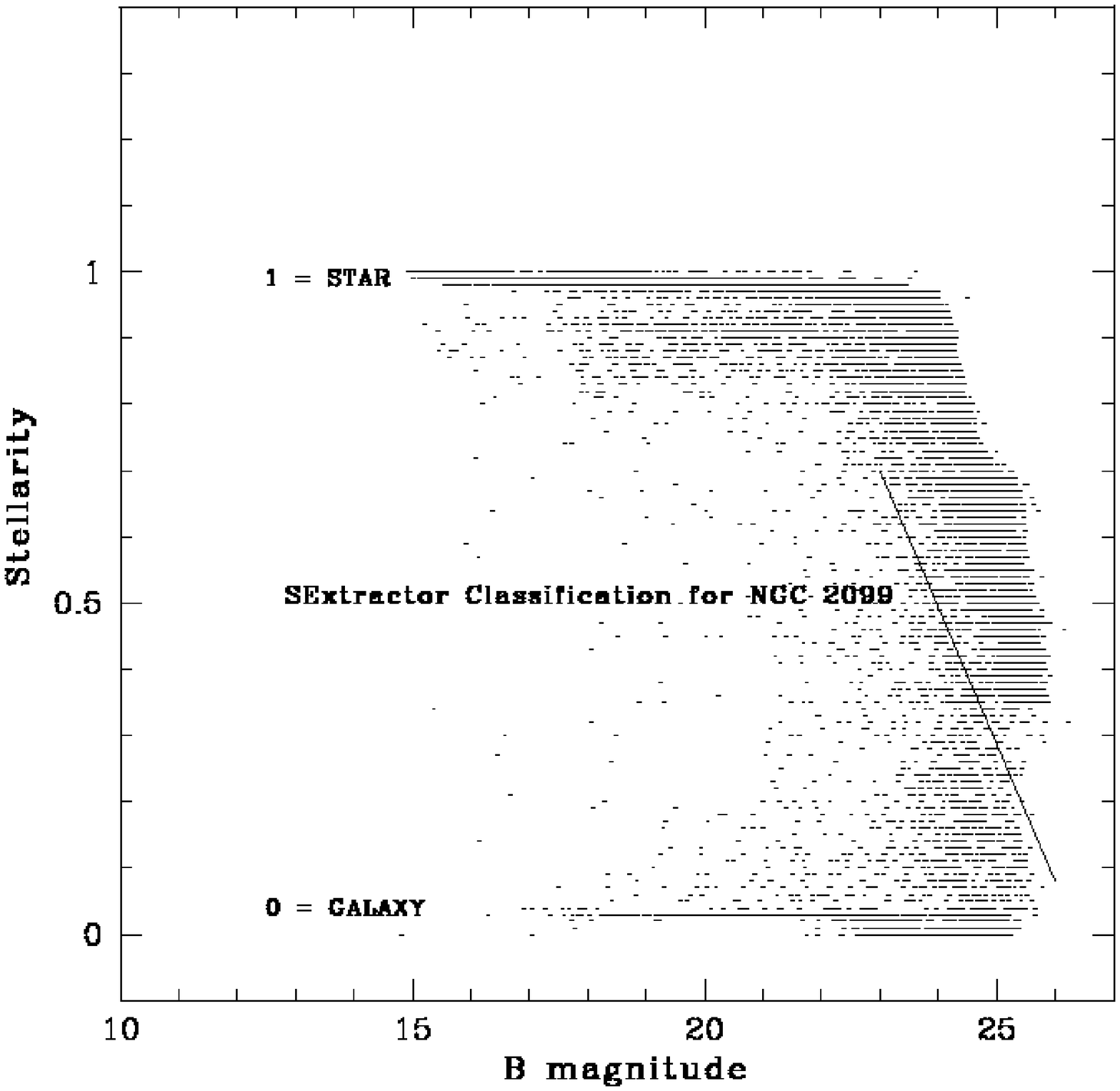]{The star/galaxy classification from
SExtractor seems to indicate two sequences (split by the line),
where the top sequence may be indicative of stars and the bottom
of galaxies. \label{N2099stellb}}

\plotone{Kalirai1.fig1.eps}

\clearpage

\figcaption[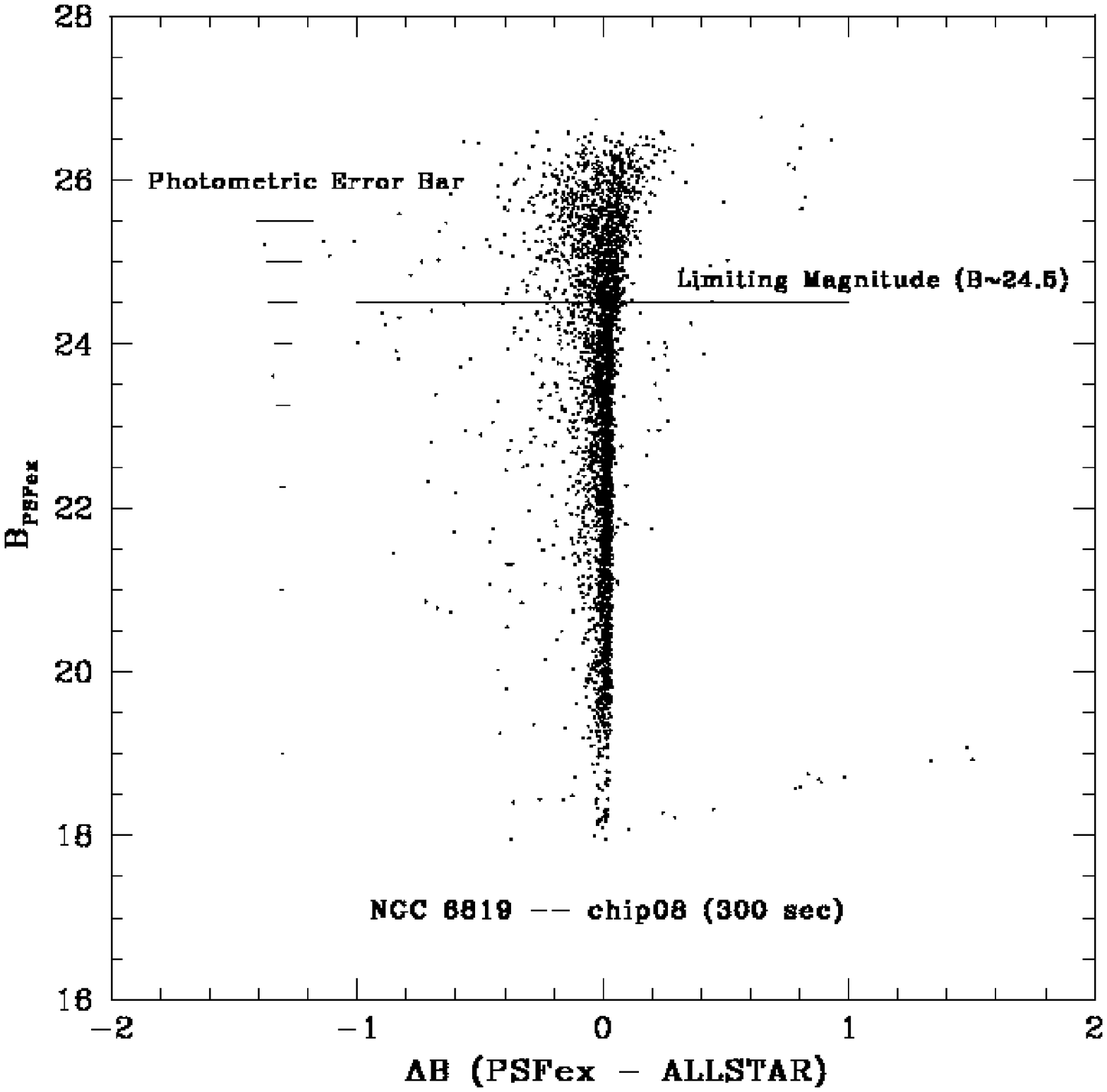]{A comparison of PSFex magnitudes
with ALLSTAR magnitudes indicates good agreement between the two
packages within the errors.  A very small bias is seen at faint
magnitudes (B $\geq$ 25) which is fainter than our limiting
magnitude cut-off. The bias effect on other CCDs and in other
filters was less than that shown here. \label{comp}}

\plotone{Kalirai1.fig2.eps}

\clearpage

\figcaption[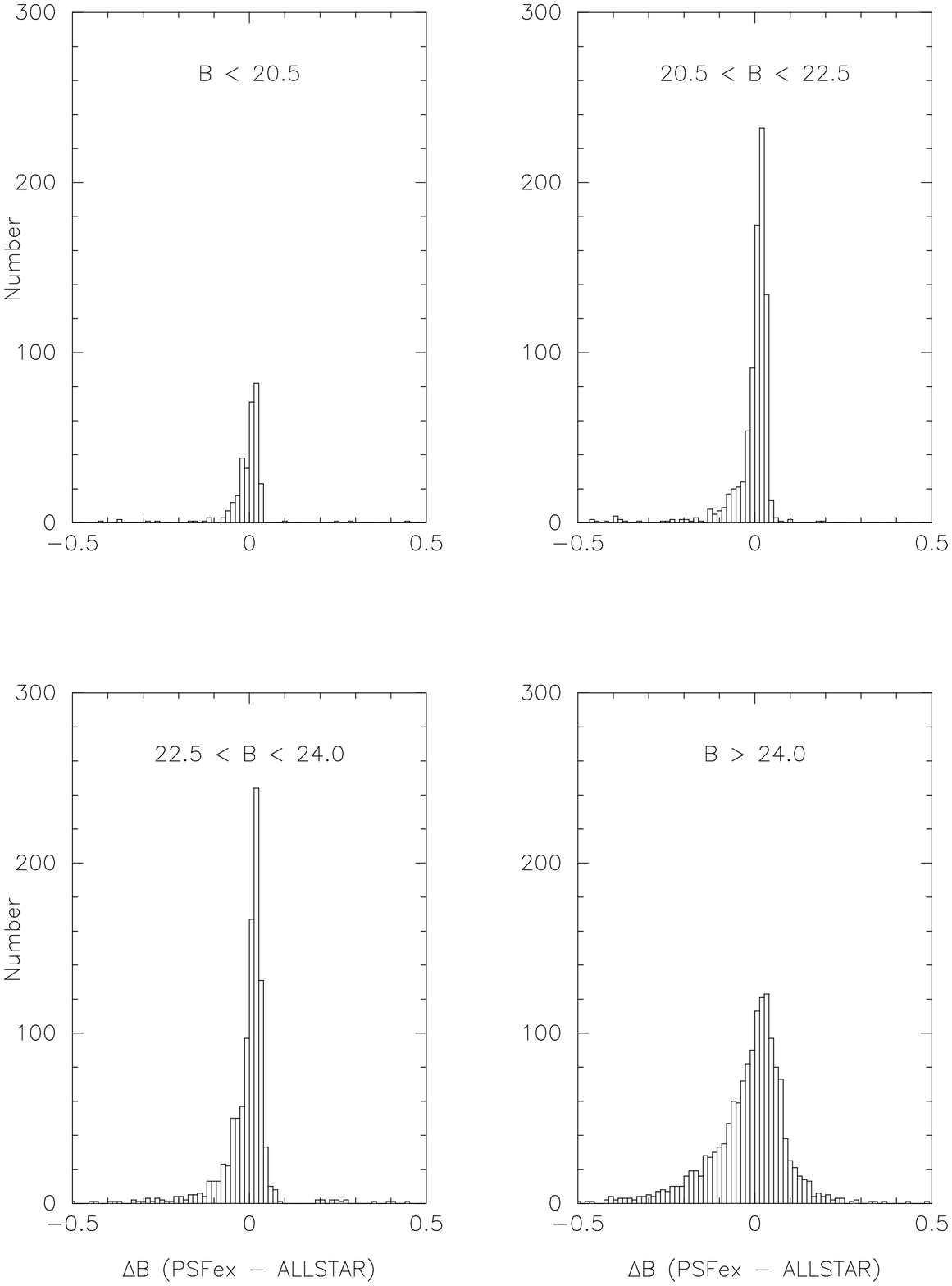]{Histograms for different magnitude
cuts show a large pile up of stars with $\Delta$B = 0 in the PSFex - 
ALLSTAR plane.  The spread in the data is small for all but very 
faint objects.  \label{histogram}}

\epsscale{0.80}
\plotone{Kalirai1.fig3.eps}

\clearpage

\figcaption[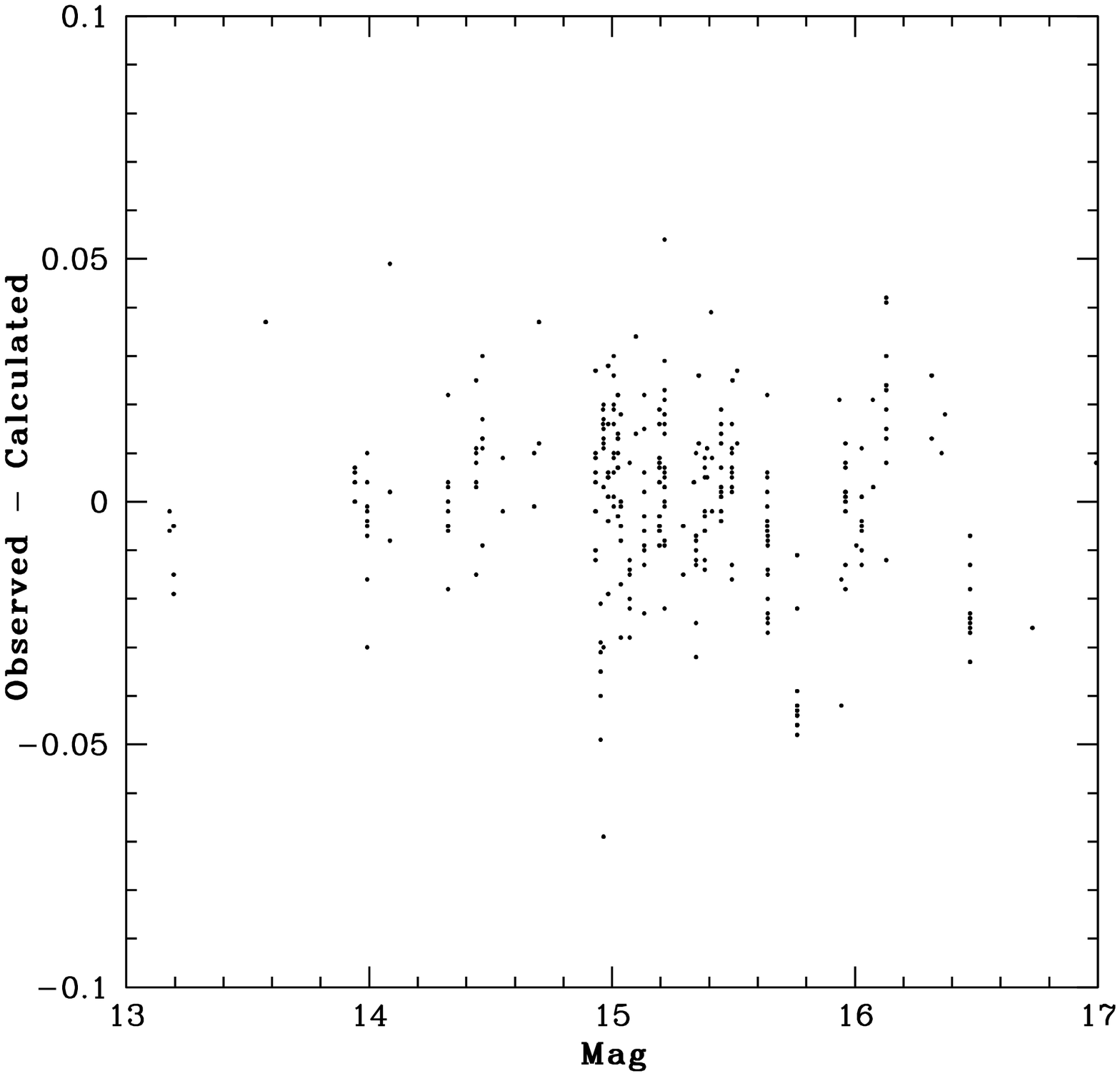]{An O-C diagram for calibration
stars on all three nights, and in both the V and B filters shows
no biasses or trends.  The small vertical spread is caused by
slightly different measurements of common stars at different
air-masses and/or exposure times. \label{calibfigure}}

\plotone{Kalirai1.fig4.eps}

%% No more than seven \figcaption commands are allowed per page,
%% so if you have more than seven captions, insert a \clearpage

%% There must be a \figcaption command for each legend. Key the text of the
%% legend and the optional \label in curly braces. If you wish, you may
%% include the name of the corresponding figure file in square brackets.
%% The label is for identification purposes only. It will not insert the
%% figures themselves into the document.
%% If you want to include your art in the paper, use \plotone.
%% Refer to the on-line documentation for details.

%% Tables should be submitted one per page, so put a \clearpage before

%% Two options are available to the author for producing tables:  the
%% deluxetable environment provided by the AASTeX package or the LaTeX
%% table environment.  Use of deluxetable is preferred.
%%

%% Three table samples follow, two marked up in the deluxetable environment,
%% one marked up as a LaTeX table.

%% In this first example, note that the \tabletypesize{}
%% command has been used to reduce the font size of the table.
%% Note also that the \label command needs to be placed

\clearpage

\begin{deluxetable}{ccccccccc}
\tabletypesize{\scriptsize} \tablecaption{The CFHT Open Star
Clusters \label{table1}} \tablewidth{0pt}
\tablehead{\colhead{Cluster} & \colhead{$\alpha$ (1950)} &
\colhead{$\delta$ (1950)} & \colhead{Night} &
\colhead{Log(Age)$^{1}$} & \colhead{(m$-$M)$_{V}^{1}$} &
\colhead{E(B$-$V)$^{1}$} & \colhead{Angular Size $(')^{1}$} &
\colhead{References}} \startdata

NGC 1039 (M34) & 2 38.8 & 42 34 & 1 & 8.26 & 8.82 & 0.10 & 35
&1,2\\

NGC 2323 (M50) & 7 00.8 & -8 16 & 1 & 8.05 & 10.73 & 0.23 & 16
&3,4
\\

NGC 6633 & 18 25.3 & 6 32 & 1 & 8.66 & 8.43 & 0.16 & 27 &5,6,7\\

NGC 1342 & 3 28.4 & 37 10 & 1 & 8.71 & 9.46 & 0.26 & 14 & \nodata\\

NGC 2343  & 7 05.9 & -10 34 & 1 & 7.36 & 10.19 & 0.15 & 7 &8,9\\

NGC 6819 & 19 39.6 & 40 04 & 1 & 9.3 & 12.3 & 0.16 & 8 &10-14\\

IC 4665 & 17 43.8 & 5 44 & 2 & 7.58 & 8.29 & 0.19 & $\geq$ 30
&15-20\\

NGC 1778 & 5 04.7 & 36 59 & 2 & 8.05 & 11.93 & 0.34 & 7 &21,22\\

NGC 225 & 0 40.5 & 61 31 & 2 & 8.23 & 9.74 & 0.24 & 12 &23\\

STOCK 2 & 2 11.4 & 59 02 & 2 & 8.23 & 8.62 & 0.38 & $\geq$ 30
&24\\

NGC 1647 & 4 43.1 & 18 59 & 2 & 8.22 & 9.85 & 0.41 & 45 &25\\

NGC 1960 (M36) & 5 32.8 & 34 06 & 2 & 7.5 & 11.25 & 0.22 & 12
&26\\

NGC 2301 & 6 49.2 & 0 31 & 2 & 8.19 & 9.79 & 0.04 & 12 &27,28\\

NGC 1750 & 5 00.9 & 23 35 & 2 & 8.30 & 10.08 & 0.34 & $\geq$ 20
&29,30\\

NGC 2099 (M37) & 5 49.1 & 32 32 & 2 & 8.6 & 11.8 & 0.28 & 26
&31-34
\\

NGC 7243 & 22 13.3 & 49 38 & 2 & 7.9 & 10.11 & 0.24 & 21 &35\\

NGC 2169 & 6 05.6 & 13 58 & 3 & 7.18 & 10.67 & 0.19 & 7 &36-40\\

NGC 2251 & 6 32.0 & 8 24 & 3 & 8.31 & 11.26 & 0.22 & 10 & \nodata\\

NGC 2168 (M35) & 6 05.8 & 24 21 & 3 & 8.05 & 10.44 & 0.25 & 28
&41,42\\

\enddata

%% Text for table notes should follow after the \enddata but before
%% the \end{deluxetable}. Make sure there is at least one \tablenotemark
%% in the table for each \tablenotetext.

\tablenotetext{1}{These parameters are not well constrained.}

%\tablecomments{}

\tablerefs{(1) Jones \& Prosser 1996; (2) Ianna \& Schlemmer 1993;
(3) Claria, Piatti, \& Lapasset 1998; (4) Schneider 1987; (5)
Jeffries 1997; (6) Reimers \& Koester 1994; (7) Sanders 1973; (8)
Maitzen 1993; (9) Claria 1972; (10) Rosvick \& VandenBerg 1998;
(11) Auner 1974; (12) Lindoff 1972; (13) Sanders 1972; (14)
Burkhead 1971; (15) Menzies \& Marang 1996; (16) Prosser \&
Giampapa 1994; (17) Prosser 1993; (18) Sanders \& van Altena 1972;
(19) Abt, Bolton \& Levy, S.G. 1972; (20) Hogg \& Kron 1955; (21)
Garcia-Pelayo \& Alfaro 1984; (22) Barbon \& Hassan 1974; (23)
Lattanzi, Massone \& Munari 1991; (24) Foster \it et al.
\normalfont 2000; (25) Turner 1992; (26) Sanner \it et al.
\normalfont 2000; (27) Marie 1992; (28) Harrington 1992; (29) Tian
\it et al. \normalfont 1998; (30) Galadi-Enriquez, Jordi, Trullols
1998; (31) Mermilliod \it et al. \normalfont 1996; (32) Becker \&
Svolopoulos 1976; (33) West 1967; (34) Upgren 1966; (35) Hill \&
Barnes 1971; (36) Pena \& Peniche 1996; (37) Peria \& Peniche
1994; (38) Perry, Lee \& Barnes 1978; (39) Sagar 1976; (40) Cuffey
\& McCuskey 1956; (41) Sung \& Bessell 1999, (42) Cudworth 1972.}

%% If you use the table environment, please indicate horizontal rules using
%% Do not put multiple tabular environments within a single table.
%% The optional \label should appear inside the \caption command.

\end{deluxetable}

\clearpage

\begin{deluxetable}{ccccccccc}
\tabletypesize{\scriptsize} \tablecaption{Observational Data for
CFHT Open Star Clusters \label{table2}} \tablewidth{0pt}
\tablehead{\colhead{Cluster} & \colhead{V (300/50/10)$^{1}$} &
\colhead{B (300/50/10)$^{1}$} & \colhead{R (50/10)$^{1}$} &
\colhead{Seeing ($''$)$^{2}$ (V/B/R)} & \colhead{Air-mass X}}
\startdata

NGC 1039$^{3}$ (M34) & 1/1/1 & 1/1/1 & 1/1 & $\sim$(0.8/0.75/0.65)
& 1.18 \\

NGC 2323$^{4}$ (M50) & 1/1/1 & 1/1/1 & 1/1 &
$\sim$(0.85/0.95/0.76) & 1.15
\\

NGC 6633 & 1/1/1 & 1/1/1 & 1/1 & $\sim$(0.6/0.75/0.60) & 1.2
\\

NGC 1342 & 9/1/1 & 9/1/1 & 1/1 & $\sim$(0.81/0.72/0.55) & 1.09
\\

NGC 2343 & 1/1/1 & 1/1/1 & 1/1 & $\sim$(0.87/1.05/0.76) & 1.23
\\

NGC 6819$^{4,5}$ & 9/1/1 & 9/1/1 & 1/1 & $\sim$(0.70/0.89/0.66) &
1.1 - 1.65  \\

IC 4665$^{3}$ & 1/1/1 & 1/1/1 & 1/1 & $\sim$(0.88/1.1/0.96) & 1.3
- 1.8  \\

NGC 1778 & 1/1/1 & 1/1/1 & 1/1 & $\sim$(0.76/0.87/0.74) & 1.15
\\

NGC 225 & 1/1/1 & 1/1/1 & 1/1 & $\sim$(0.82/0.92/0.77) & 1.37
\\

STOCK 2$^{3}$ & 1/1/1 & 1/1/1 & 1/1 & $\sim$(0.78/0.82/0.75) &
1.36  \\

NGC 1647$^{3}$ & 1/1/1 & 1/1/1 & 1/1 & $\sim$(0.65/0.77/0.63) &
1.23  \\

NGC 1960 (M36) & 1/1/1 & 1/1/1 & 1/1 & $\sim$(0.80/0.82/0.67) &
1.05  \\

NGC 2301 & 1/1/1 & 1/1/1 & 1/1 & $\sim$(0.85/0.88/0.75) & 1.37
\\

NGC 1750 & 1/1/1 & 1/1/1 & 1/1 & $\sim$(0.77/0.93/0.85) & 1.20
\\

NGC 2099$^{4,6}$ (M37) & 3/1/1 & 3/1/1 & 1/1 &
$\sim$(0.90/0.82/0.81) & 1.03  \\

NGC 7243$^{3}$ & 1/1/1 & 1/1/1 & 1/1 & $\sim$(0.78/0.79/0.80) &
1.16  \\

NGC 2169$^{2}$ & 2/1/1 & 2/1/2 & 2/1 & $\sim$(1.8/1.8/1.6) & 1.06
\\

NGC 2251$^{2}$ & 2/1/1 & 2/1/1 & 1/1 & $\sim$(1.35/1.30/1.75) &
1.37
\\

NGC 2168$^{4}$ (M35) & 2/1/2 & 2/1/2 & 1/1 &
$\sim$(1.35/1.20/0.90) & 1.60 \\

\\
\tableline

Field & V (5/10)$^{1,7}$ & B (5/10/15/20)$^{1,7}$ & R
(5/10)$^{1,7}$
\\

\tableline

CALIBRATION SA-92 & 1/2 & 2/0/2/0 & 1/2 \\

CALIBRATION SA-92 & 2/2 & 2/2/0/1 & 2/2  \\

CALIBRATION SA-92 & 0/2 & 0/0/1/0 & 0/2  \\

CALIBRATION SA-95 & 0/2 & 0/2/0/0 & 0/2 \\
\\
\tableline

Frame & V Flat Field & B Flat Field & R Flat Field & Bias (Dark
Filter) \\ \tableline

PRE-PROCESSING & 33 & 20 & 9 & 13 \\
\enddata

%% Text for table notes should follow after the \enddata but before
%% the \end{deluxetable}. Make sure there is at least one \tablenotemark
%% in the table for each \tablenotetext.

\tablenotetext{1}{The number of exposures taken in this filter,
for each exposure time (sec) listed.}

\tablenotetext{2}{The data for these clusters suffer from one or
more of the following : high humidity, poor focus, or cirrus.}

\tablenotetext{3}{An equivalent number of blank field images were
also obtained for these clusters (offset by 1 degree).}

\tablenotetext{4}{Further reductions are being completed for these
clusters.  To be published shortly.}

\tablenotetext{5}{Additional 0.2 second and 1.0 second images were
obtained at a later date.}

\tablenotetext{6}{Additional 0.5 second images were obtained at a
later date.}

\tablenotetext{7}{These images are intentionally taken over a
range of air-masses to obtain air-mass terms.}

%\tablecomments{}

\end{deluxetable}

\clearpage

\begin{deluxetable}{cccccccccccccc}
\tabletypesize{\scriptsize} \tablecaption{Calibration Stars For
the 12 CCDs (00-12) \label{table4}} \tablewidth{0pt}
\tablehead{\colhead{Exposure$^1$} & \colhead{Air-mass X} &
\colhead{00} & \colhead{01} & \colhead{02} & \colhead{03} &
\colhead{04} & \colhead{05} & \colhead{06} & \colhead{07} &
\colhead{08} & \colhead{09} & \colhead{10} & \colhead{11}}
\startdata \bf NIGHT 1 -- SA-92 \rm &
\\

V/5 & 1.085 & 0 & 0 & 3 & 1 & 5 & 0 & 1 & 6 & 2 & 1 & 0 & 0 \\

B/5 & 1.101 & 0 & 0 & 3 & 1 & 5 & 0 & 1 & 6 & 3 & 2 & 0 & 0 \\

B/5 & 1.099 & 0 & 0 & 3 & 1 & 5 & 0 & 1 & 6 & 2 & 1 & 0 & 0 \\

V/10 & 1.088 & 0 & 0 & 3 & 1 & 4 & 0 & 1 & 5 & 2 & 0 & 0 & 0 \\

V/10 & 2.105 & 0 & 0 & 3 & 1 & 5 & 0 & 1 & 6 & 2 & 1 & 0 & 0 \\

B/15 & 1.092 & 0 & 0 & 3 & 1 & 5 & 0 & 1 & 6 & 2 & 0 & 0 & 0 \\

B/15 & 2.038 & 0 & 0 & 3 & 1 & 5 & 0 & 1 & 6 & 2 & 0 & 0 & 0 \\

\bf NIGHT 2 -- SA-92 \rm &
\\

B/5 & 1.519 & 0 & 0 & 3 & 1 & 5 & 0 & 1 & 6 & 3 & 1 & 0 & 0 \\

B/5 & 1.060 & 0 & 0 & 3 & 1 & 5 & 0 & 1 & 6 & 0 & 1 & 0 & 0 \\

B/10 & 1.508 & 0 & 0 & 3 & 1 & 5 & 0 & 1 & 6 & 2 & 1 & 0 & 0 \\

B/10 & 1.060 & 0 & 0 & 3 & 1 & 5 & 0 & 1 & 6 & 2 & 1 & 0 & 0 \\

B/20    & 1.496 & 0 & 0 & 3 & 1 & 5 & 0 & 1 & 6 & 2 & 1 & 0 & 0 \\

V/5  & 1.549 & 0 & 0 & 3 & 1 & 5 & 0 & 1 & 6 & 2 & 1 & 0 & 0 \\

V/5  & 1.059 & 0 & 0 & 3 & 1 & 4 & 0 & 1 & 5 & 2 & 1 & 0 & 0 \\

V/10 & 1.535 & 0 & 0 & 3 & 1 & 5 & 0 & 1 & 5 & 2 & 1 & 0 & 0 \\

V/10 & 1.059 & 0 & 0 & 3 & 1 & 4 & 0 & 0 & 4 & 2 & 1 & 0 & 0 \\

\bf NIGHT 2 -- SA-95 \rm &
\\

B/10 & 1.389 & 1 & 0 & 0 & 0 & 0 & 0 & 3 & 2 & 4 & 0 & 2 & 0 \\

B/10 & 1.398 & 0 & 0 & 0 & 0 & 0 & 0 & 3 & 2 & 4 & 0 & 2 & 0 \\

V/10 & 1.413 & 1 & 0 & 0 & 0 & 0 & 0 & 3 & 2 & 4 & 0 & 0 & 0 \\

V/10 & 1.422 & 1 & 0 & 0 & 0 & 0 & 0 & 3 & 3 & 4 & 0 & 1 & 0 \\

\bf NIGHT 3$^{2}$ -- SA-92 \rm &
\\

V/10 & 1.076 & 0 & 0 & 8 & 1 & 5 & 0 & 5 & 8 & 2 & 1 & 0 & 0 \\

V/10 & 1.078 & 0 & 0 & 8 & 1 & 5 & 0 & 5 & 8 & 2 & 1 & 0 & 0 \\

B/15  & 1.081 & 0 & 0 & 8 & 1 & 8 & 0 & 6 & 9 & 2 & 1 & 0 & 0 \\

\enddata

%% Text for table notes should follow after the \enddata but before
%% the \end{deluxetable}. Make sure there is at least one \tablenotemark
%% in the table for each \tablenotetext.

\tablenotetext{1}{Column organized as filter/exposure time
(seconds).}

\tablenotetext{2}{We used night 1 calibration results to create
additional stars for this night.}

%\tablecomments{}

\end{deluxetable}

%% You can append references to a table using the \tablerefs command.

%% Tables may also be prepared as separate files. See the accompanying
%% sample file table.tex for an example of an external table file.
%% To include an external file in your main document, use the \input
%% command. Uncomment the line below to include table.tex in this
%% sample file.

%\input{table}

%% The following command ends your manuscript. LaTeX will ignore any text

\end{document}